\documentclass[conference]{IEEEtran}
 \IEEEoverridecommandlockouts

\usepackage{cite}

\onecolumn
\usepackage{amsmath}
\usepackage{amsfonts}
\usepackage{amssymb}
\usepackage{amsthm}  
\usepackage{bbm}  
\usepackage{color}
 \usepackage{amssymb}
\usepackage{txfonts}
\usepackage[T1]{fontenc}

\newtheorem{thm}{Theorem}

\newtheorem{Lemma}{Lemma}

\theoremstyle{remark}
\newtheorem{Remark}{Remark}

\newtheorem{Definition}{Definition}

\ifCLASSINFOpdf
   \usepackage[pdftex]{graphicx}
\else
\fi
\usepackage[scr=dutchcal]{mathalfa}
\let\mathscr\mathbscr

\usepackage{makecell}
\usepackage{multirow}
\usepackage{tikz}
\usepackage[scr=dutchcal]{mathalfa}
\let\mathscr\mathbscr

\newcolumntype{x}[1]{>{\centering\arraybackslash}p{#1}}

\newcommand\diag[4]{%
  \multicolumn{1}{p{#2}|}{\hskip-\tabcolsep
  $\vcenter{\begin{tikzpicture}[baseline=0,anchor=south west,inner sep=#1]
  \path[use as bounding box] (0,0) rectangle (#2+2\tabcolsep,\baselineskip);
  \node[minimum width={#2+2\tabcolsep-\pgflinewidth},
        minimum  height=\baselineskip+\extrarowheight-\pgflinewidth] (box) {};
  \draw[line cap=round] (box.north west) -- (box.south east);
  \node[anchor=south west] at (box.south west) {#3};
  \node[anchor=north east] at (box.north east) {#4};
 \end{tikzpicture}}$\hskip-\tabcolsep}}

\begin{document}
%
\title{Quasi Linear Codes: Application to Point-to-Point and Multi-Terminal Source Coding}

\author{Farhad Shirani
  \IEEEauthorblockN{}
  \IEEEauthorblockA{EECS Department\\University of Michigan\\ Ann Arbor,USA \\
    Email: fshirani@umich.edu } 
\and
 \IEEEauthorblockN{Mohsen Heidari}
  \IEEEauthorblockA{EECS Department\\University of Michigan\\ Ann Arbor,USA \\
    Email: mohsenhd@umich.edu} 
 
  \and
  \IEEEauthorblockN{S. Sandeep Pradhan}
  \IEEEauthorblockA{EECS Department\\University of Michigan\\ Ann Arbor,USA \\
    Email: pradhanv@umich.edu}
}


%


\maketitle

\begin{abstract}
 A new ensemble of structured codes is introduced. These codes are called Quasi Linear Codes (QLC). 
The QLC's are constructed by taking subsets of linear codes. They have a looser structure 
compared to linear codes and are not closed under addition. We argue that these codes provide gains in terms of 
achievable Rate-Distortions (RD) in different multi-terminal source coding problems.  We derive the necessary 
covering bounds for analyzing the performance of QLC's. We then consider the Multiple-Descriptions (MD) problem, and prove through an example that 
the application of QLC's gives an improved achievable RD region for this problem. 
Finally, we derive an inner bound to the achievable RD region for the general MD problem which strictly 
contains all of the previous known achievable regions.


\end{abstract}

\section{Introduction}

\IEEEPARstart{E}{xploiting} algebraic structure in multi-terminal communication problems has been of recent interest. 
Structured coding provides significant improvements in terms of asymptotic performance for many such problems   
\cite{KM,BC,MA,CJ,IC,MD}. Korner and Marton first observed the phenomenon in a distributed source 
coding (DSC) problem involving  the compression of a modulo-two addition of two correlated 
binary sources \cite{KM}. Such gains have also been observed in other multi-terminal  problems such as 
multiple-access channel with states available at the transmitters \cite{Zamir}, 
computation over multiple-access channels \cite{Nazer}, the interference channel \cite{IC}, 
the broadcast channel \cite{BC}, and the MD source coding \cite{MD}.
In the large body of work dedicated to this topic various types of structured codes have been 
considered. The most well-studied of these codes are linear codes. These codes are constructed 
over finite fields and are closed with respect to the linear operation associated with the field. 

Korner and Marton suggest the use of identical linear codes to effect binning of two correlated binary sources 
when the objective is to reconstruct the modulo-two sum of the sources at the decoder. They show that such an approach 
leads to optimality. However, if the objective is to 
have the complete reconstruction of  both the sources at the decoder (Slepian-Wolf setting), 
then it has been shown that 
for certain sources, using identical binning can be strictly suboptimal \cite{sandeep}. In general, to achieve the 
Slepian-Wolf performance limit, one needs to use either binning of the two sources using two independent 
linear codes or use independent unstructured binning of the two sources. Moreover, there is no 
known method based on unstructured codes which achieves optimality for the reconstruction of the modulo-two sum.
In summary, the former requires only identical binning, where as the latter requires only independent binning.
  
This leads to the following question: (i) is there a spectrum of strategies involving 
partially independent binning of the two sources that lie between these two extremes, and (ii) is there a class 
of problems for which such strategies give gains in asymptotic performance? 
In other words, is there a \textit{trade-off}  between structured coding and unstructured coding. 
Based on this intuition, in this paper, we consider codes which are not fully closed 
under any algebraic structure but maintain a degree of \textit{"closedness"} with respect to some.   

We introduce a new class of structured code ensembles called QLC whose 
closedness can be controlled. A QLC is a  subset  of a linear code. It is difficult to 
analyze the performance of arbitrary subsets of linear codes. Instead, we provide a method for 
constructing specific subsets of these codes by putting single-letter distributions on the indices of the 
codewords. We analyze the performance of the resulting ensemble. We are able to characterize the asymptotic 
performance using single-letter information quantities. By choosing the single-letter distribution on the 
indices one can operate anywhere in the spectrum between the two extremes: linear codes and unstructured codes. 
First, we show that QLC's achieve the Shannon rate-distortion function for discrete memoryless sources with 
bounded additive distortion functions. Next, we  show through an MD
source coding example that  application of QLC's gives a better inner bound to 
the optimal achievable RD region 
compared to the best known MD coding strategies. Finally, we provide a new inner bound to the optimal 
achievable RD region for the general MD problem. The method builds upon the Sperner Set Coding (SSC) 
scheme introduced in \cite{MD}.

The rest of the paper is organized as follows: Section \ref{sec:not} provides the notation 
used in this paper. Section \ref{sec:def} defines the new coding structures. In section \ref{sec:trade},
 we investigate the tradeoff mentioned above. Section \ref{sec:MD} includes an 
example in the three-descriptions problem where QLC's give improvements. 
Also, we give a new inner bound for the achievable RD region for 
the general MD problem. Section \ref{sec:conc} concludes the paper.

\section{Notation}
\label{sec:not}
Random variables are denoted by capital letters such as $X, U$, their realizations by small letters $x,u$, and their corresponding alphabets (finite) by sans-serif typeface $\mathsf{X}$, $\mathsf{U}$, respectively. Numbers are represented by small letters such as $l, k$. We denote the the field of size $q$ by $\mathbb{F}_q$. The set of numbers $\{1,2,\ldots,m\}$ is represented by $[1,m]$. Vectors are represented by the bold type-face such as $\mathbf{u}, \mathbf{b}$. For a random variable $X$, $A_{\epsilon}^n(X)$ denotes the set of $\epsilon$-typical sequence of length $n$ with respect to $P_X$, where we use the definition of frequency typicality. Let $q$ be a prime number. For $l\in\mathbb{N}$, consider $U_i,i\in[1,m]$ i.i.d random variables with distribution $P_U$ defined on a field $\mathbb{F}_q$. $U^{\otimes{l}}$ denotes a random variable which has the same distribution as $\sum_{i\in [1,l]}U_i$ where the summation is over $\mathbb{F}_q$.
\section{New Codebook Structures}
\label{sec:def}
In this section, we define our new coding structures and provide the foundations for their analysis. 
\subsection{Quasi Linear Codes}
First, we define a new ensemble of codes called QLC's.  
The ensemble is defined over a finite field $\mathbb{F}_q$ where $q$ is a prime number. The codebooks are constructed by first generating the coset of a linear code called a coset code. 
\begin{Definition} 
 A  $(k,n)$ coset code $\mathcal{C}$ is characterized by a generator matrix $G_{k\times n}$ and a dither $\mathbf{b}^n$ defined on the field $\mathbb{F}_q$. $\mathcal{C}$ is defined as follows:
\begin{equation*}
\mathcal{C}\triangleq\{\mathbf{u}G+\mathbf{b}|\mathbf{u}\in \mathbb{F}_q^k\}. 
\end{equation*}
The rate of the code is given by $R=\frac{k}{n}\log{q}$.
\end{Definition}
A QLC is a subset of a linear code, the following provides the definition of a QLC:
\begin{Definition} 
\label{def:Qlingen}

A $(k,n)$ QLC is characterized by a generator matrix $G_{k\times n}$, a dither $\mathbf{b}^n$ and a set $\mathsf{U}$ defined on $\mathbb{F}_q$. The codebook is defined as follows:
\begin{equation*}
\mathcal{C}\triangleq\{\mathbf{u}G+\mathbf{b}|\mathbf{u}\in \mathsf{U}\}. 
\end{equation*}
If $G$ is injective on $\mathsf{U}$, then the rate of the code is given by $R=\frac{1}{n}\log{|\mathcal{C}|}=\frac{1}{n}\log{|\mathsf{U}|}$.
\end{Definition}
It is difficult to analyze the performance of such codes for general sets $\mathsf{U}$. In this paper, we focus on the case when $\mathsf{U}$ is a cartesian product of typical sets. More precisely, let $m\in \mathbb{N}$, $\epsilon \in \mathbb{R}^+$, and $U_1, U_2,..., U_m$ be random variables defined on $\mathbb{F}_q$. Consider natural numbers $k_i, i\in [1,m]$ such that $\sum_{i\in [1,m]}k_i=k$. Construct generator matrices $G_i$ with dimension $k_i\times n$. We are interested in analyzing the performance of codebooks of the following form:
\begin{equation*}
\mathcal{C}\triangleq \{\sum_{i\in [1,m]}\mathbf{u}_iG_i+\mathbf{b}|\mathbf{u}_i\in A_{\epsilon}^{k_i}(U_i)\}. 
\end{equation*}
In this case, the rate of the code is $R=\sum_{i\in[1,m]}\frac{1}{n}\log|A_\epsilon^{k_i}(U_i)|$ which approaches $\sum_{i\in[1,m]}\frac{k_i}{n}H(U_i)$ as $n\to \infty, \epsilon\to 0$.
\begin{Remark}
 In the notation of Definition \ref{def:Qlingen}, $G=[G_1^t|G_2^t|...|G_m^t]^t$ and $\mathbf{u}=(\mathbf{u}_1,\mathbf{u}_2,...,\mathbf{u}_m)$.
\end{Remark}
\begin{Remark}
While we concentrate on the case when $\mathsf{U}= \bigotimes_{i\in[1,m]}A_\epsilon^{k_i}(U_i)$, it is possible to carry out performance analysis of such an ensemble of codebooks when $\mathsf{U}$ is taken to be more general. For example, a more general result can be obtained by taking $\mathsf{U}$ to be a joint typical set of vectors of correlated random variables $U_1,U_2,...,U_m$. 
\end{Remark}
\begin{Remark}
 A $(k,n)$ linear code is only defined for $k\leq n$. When constructing a QLC, we take $R\leq\log{q}$. This ensures that for a randomly and uniformly generated matrix $G$, the resulting mapping is injective on $\mathsf{U}$ with high probability. However, there is no additional restrictions on the $k_i$'s. As an example, let $m=1$, one can take $k_1>n$ and $U_1$ such that $\frac{k_1}{n}H(U_1)<\log{q}$. In this case $\{\mathbf{u}_1G_1+B|\mathbf{u}_1\in A^{k_1}_\epsilon(U_1)\}$ is a codebook whose rate is close to $\frac{k_1}{n}H(U_1)$ for large $n$. Note that $G_1$ is not injective on the vector space $\mathbb{F}_q^{k_1}$.
\end{Remark}
\subsection{Nested Quasi Linear Codes}
In this section, we define Nested Quasi Linear Codes (NQLC). The following gives the definition for a pair of Nested Linear Codes (NLC): 

\begin{Definition}
\label{def:PNLC}
For natural numbers $k_i<k_o,k'_o<n$, let $G_{k_i\times n}, \Delta{G}_{(k_o-k_i)\times n}$ and $\Delta{G'}_{(k'_o-k_i)\times n}$ be matrices on $\mathbb{F}_q$. Define $\mathcal{C}_i, \mathcal{C}_o$ and $\mathcal{C}'_o$ as the linear codes generated by $G$, $[G|\Delta{G}]$ and $[G|\Delta{G}']$, respectively.  $\mathcal{C}_o$ and $\mathcal{C}'_o$ are called a pair of NLC's with inner code $\mathcal{C}_i$. We denote the outer rates as $r_o=\frac{k_o}{n}$ and $r'_o=\frac{k'_o}{n}$, and the inner rate $r_i=\frac{k_i}{n}$.
  \end{Definition}
A pair of NQLC's are defined as follows: 
\begin{Definition}
For natural numbers $k_1,k_2,\dotsb, k_m$, let $G_{k_i\times n}, i\in [1,m]$ be matrices on $\mathbb{F}_q$, and let $\mathbf{b_j},j \in\{1,2\}$ be two dithers on the field. Also, let $(U_1,U_2,\dotsb, U_m)$ and $(U'_1,U'_2,\dotsb, U'_m)$ be a pair of random vectors on $\mathbb{F}_q$. The pair of QLC's characterized by the matrices $G_{k_i\times n}, i\in [1,m]$, and each of the two vectors of random variables and dithers are called a pair of NQLC's.\label{def:NQLC}
  \end{Definition}

The definition of the NQLC's is a generalization of NLC's. To see this, consider an arbitrary pair of NLC's with the parameters as in Definition \ref{def:PNLC}. These two codes are a pair of NQLC's with parameters $m=3$, $U_1, U_2$ and $U'_1,U'_3$ uniform, $U_3$ and $U'_2$  constants and $k_1=k'_1=k_i$ and $k_2=k_o-k_i, k'_3=k'_o-k_i$. It was shown in \cite{MD} that in the general MD problem, it is beneficial to use m-tuples of NLC's called an ensemble of NLC's. The following gives the definition for an ensemble of NLC's:
\begin{Definition}
 A set of $l$ linear codes $C^n_{k}, k\in [1,l]$ is called an ensemble of nested linear codes with parameter $(r_\mathsf{J})_{\mathsf{J}\subset[1,l]}$ if the size of the intersection $\bigcap_{k\in \mathsf{J}} C_{k}$ is equal to $2^{nr_\mathsf{J}}$ for all $\mathsf{J}\subset \mathsf{M}$.
 \end{Definition}
From the above discussions we can define an ensemble of NQLC's as follows:
\begin{Definition}
 Let $l\in \mathbb{N}$. For natural numbers $k_1,k_2,\dotsb, k_m$, let $G_{k_i\times n}, i\in [1,m]$ be matrices on $\mathbb{F}_q$, and $\mathbf{b}_j, j\in [1,l]$ dithers on the field. Also, let $(U_{i,1},U_{i,2},\dotsb, U_{i,m}), i\in [1,l]$ be vectors of random variables on $\mathbb{F}_q$. The ensemble of QLC's characterized by the matrices $G_{k_i\times n}, i\in [1,m]$ and each of the vectors of random variables and the dithers is called an ensemble of NQLC's.
\end{Definition}
Once more it is straightforward to check that this is a generalization of the definition for ensembles of NLC's. Consequently, any achievability results derived using NLC's can be obtained via NQLC's as well. As an example, the next lemma proves that combined with binning, application of these codes can achieve Shannon's RD function for PtP communication. 
\begin{Lemma}
 NQLC's achieve Shannon's RD function for PtP source coding for arbitrary source distributions and bounded additive distortion functions.
\end{Lemma}
\begin{IEEEproof}
 We provide an outline of the proof for an arbitrary source $X$ defined on $\mathbb{F}_q$. Let $p(y|x)$ be an optimizing test channel for Shannon's RD function. Take $m=1$, and $R_o=\frac{k_1}{n_1}=\log{q}-H(Y|X)$. Construct a QLC with these parameters. Bin the code randomly and uniformly with rate $\log{q}-H(Y)$. For each source sequence $x^n$, the encoder finds a codeword typical with $x^n$. The encoder transmits the bin index. The decoder finds the unique codeword in the bin which is typical with respect to $p(y)$. It is straightforward to check that with the above rates transmission can be carried out with probability of error going to 0. 
\end{IEEEproof}
\section{Fundamental Properties of QLC's}
\label{sec:trade}
As mentioned in the introduction, the application of NLC's gives gains in different multi-terminal source coding problems. These gains are a result of the fact that linear codes are closed under addition. More precisely, the sum of a pair of NLC's has a smaller size than that of two randomly generated unstructured codes of the same rates. As a result, for the two codebooks $\mathcal{C}_1$ and $\mathcal{C}_2$ it takes less rate to transmit $\mathcal{C}_1+\mathcal{C}_2$ if the two codes are nested linear codes. However, it has been shown that this closure property has its downsides as well. In fact, a tradeoff has been observed between using NLC's and unstructured codes in different communication setups. The drawback of using NLC's manifests itself in the derivation of the mutual covering bounds for these coding structures. It turns out that unstructured codes satisfy their covering constraint more easily (i.e. their covering bounds are satisfied for lower rates). The idea behind defining QLC's is to breach this gap between NLC's and unstructured codes. 

This section is divided into two parts. First, we analyze the addition of QLC's. We show that for the two codebooks $\mathcal{C}_1$ and $\mathcal{C}_2$, the sum $\mathcal{C}_1+\mathcal{C}_2$ has a higher rate than the sum of two linear codes of the same rate and a smaller rate than that of two unstructured codes. Then, in the second part, we derive the covering bounds associated with QLC's. In this part, we show that the covering bounds for QLC's are less strict than those for NLC's and more strict than the ones for unstructured codes. Using these two results, we can analyze the tradeoff mentioned above for the application of QLC's.  

\subsection{The Addition of QLC's}
QLC's are not linearly closed but at the same time maintain a degree of "closedness" in their structure. Notice that if we repeatedly add a QLC with itself, the resulting set of codevectors will be a subset of the linear code generated by $[G_1|G_2|\dotsb|G_m]$, where the $G_i$'s are the generator matrices for the QLC. Whereas, if a random unstructured code is added with itself repeatedly, the resulting space would converge to the whole vector space. 
In the following lemma we investigate the addition of $l$ copies of a QLC with each other:
\begin{Lemma} \label{lem:codesize}
For $R\in (0,\log{q})$, 
let $\mathcal{C}_Q$ be a QLC with parameters $m,n, k_1,k_2,...,k_m \in \mathbb{N}$, $U_i, i\in[1,m]$, matrices $G_i ,i\in [1,m]$, and dither $\mathbf{b}$, such that the code has rate R, where the $G_i$'s and $\mathbf{b}$ are generated randomly and uniformly on $\mathbb{F}_q$.
The probability of the following events goes to one as $n\to \infty$:
\begin{enumerate}
 \item{} $\frac{1}{n}\log|\sum_{i\in[1,l]}{\mathcal{C}_Q}|\stackrel{.}{=}\sum_{i\in[1,m]}\frac{k_i}{n}H(U_i^{\otimes l})$,
 \vspace{0.05in}
 \item{} $R\leq \frac{1}{n}\log|\sum_{i\in[1,l]}{\mathcal{C}_Q}| \leq \min{(\log{q},l\times R)}$ where equality is achieved on the left hand side by taking $U_i$'s to be uniform. 
 \end{enumerate}
\begin{IEEEproof}
 The proof follows standard typicality arguments and is omitted.
\end{IEEEproof}
\end{Lemma}
\begin{Remark}
As mentioned in the lemma, equality on the left hand side of condition 2 can be achieved by taking the $U_i$'s to be uniform. In this case the QLC becomes a coset code. On the right hand side, one can approach equality by taking $k_i>n$ and $U_i$ to be very low entropy random variables. Observe that if the random variable $U_i$ has low entropy, then $H(U_i^{\otimes l})\approx lH(U_i)$. 
\end{Remark}
 
\begin{Remark}
 For an arbitrary $n$-length codebook $\mathcal{C}$ with rate $R$, it is straightforward to show that $R\leq \frac{1}{n}\log|\sum_{i\in[1,l]}{\mathcal{C}}| \leq \min{(\log{q},l\times R)}$. For linear codes equality always holds on the left-hand side. For random codes, equality always holds on the right-hand side. Whereas, QLC's achieve all of the possible values allowed for $\frac{1}{n}\log|\sum_{i\in[1,l]}{\mathcal{C}}|$. 
 \end{Remark}

\subsection{Mutual Covering Bounds for NQLC's}
We proceed with deriving the mutual covering bounds for the NQLC's. The covering bounds are useful for determining inner bounds to achievable RD regions in different source coding settings. In this paper, we concentrate on the MD problem. The following gives a formal definition for a $P_{XV_1V_2}$-covering pair of codes. 
\begin{Definition}
\label{def:cov}
  Let $\mathbb{F}_q$ be a field. Consider 3 random variables $X$, $V_1$ and $V_2$, where $X$ is defined on an arbitrary finite set $\mathsf{X}$ and $V_1$ and $V_2$ are defined on $\mathbb{F}_q$. Fix a PMF $P_{X,V_1,V_2}$ on $\mathsf{X}\times\mathbb{F}_q\times \mathbb{F}_q$. A sequence of code pairs $(C_1^n,C_2^n)$ is called $P_{XV_1V_2}$-covering if:
\begin{align*} 
 &\forall \epsilon>0, P(\{x^n|\exists (v_1^n,v_2^n)\in A_{\epsilon}^n(V_1,V_2|x^n)\cap C_1\times C_2\})\to 1,\end{align*}
\label{P_cov}
 $\text{ as } n\to \infty.$
\end{Definition}
As mentioned in \cite{MD}, deriving the achievable RD region for the MD setup using the SSC scheme involves obtaining the mutual covering bounds for independently generated codebooks. The following lemma characterizes these bounds for a pair of unstructured codes.
\begin{Lemma}
\label{lem:unst}
\cite{ElGamalLec}
For any distribution $P_{XV_1V_2}$ on $\mathsf{X}\times\mathbb{F}_q\times \mathbb{F}_q$ and rates $r_1,r_2$ satisfying (\ref{unst1})-(\ref{unstend}), there exists a sequence of pairs of unstructured codes $\mathcal{C}^n_1$ and $\mathcal{C}^n_2$ which are $P_{XV_1V_2}$-covering.
\begin{align}
&r_1  \geq H(V_1) - H(V_1|X)\label{unst1}\\
&r_2 \geq H(V_2) - H(V_2|X)\\
&r_1+r_2 \geq H(V_1,V_2) -H(V_1,V_2|X)
\label{unstend}
\end{align}
\end{Lemma}
When using ensembles of NLC's, new covering bounds are necessary since the codebooks are not independently generated (e.g. they  share a common inner code.). The next lemma presents the bounds for a pair of NLC's.

\begin{Lemma}\label{lem:NLC}\cite{MD}
For any $P_{XV_1V_2}$ on $\mathsf{X}\times\mathbb{F}_q\times \mathbb{F}_q$ and rates $r_o=r_1$, $r'_o=r_2$ and $r_i$ satisfying \ref{NLC1}-\ref{NLC4}, there exists a sequence of pairs of NLC's $\mathcal{C}^n_1$ and $\mathcal{C}^n_2$ which are $P_{XV_1V_2}$-covering.
\begin{align}
&r_1  \geq \log q - H(V_1|X)\label{NLC1}\\
&r_2 \geq \log q - H(V_2|X)\label{NLC2}\\
&r_1+r_2 \geq 2\log q -H(V_1,V_2|X)\label{NLC3}\\
&r_1+r_2-r_i \geq \max_{\alpha,\beta\in \mathbb{F}_q\backslash{\{0\}}}(\log{q}- H(\alpha V_1+ \beta V_2|X)),
\label{NLC4}
\end{align}
\end{Lemma}

In the process of deriving the inner bound to the achievable RD region, the entropy terms in Lemma \ref{lem:unst} and $\log q$ terms in Lemma \ref{lem:NLC} vanish in the Fourier-Motzkin elimination and only the conditional entropy terms would remain on the RHS \cite{MD}. So, the only consequential difference between the two bounds lies in the introduction of inequality (\ref{NLC4}). First, we argue that this inequality can not be eliminated by a more precise error analysis. We use a converse coding argument to prove this point. Assume the existence of a pair of NLC's $\mathcal{C}_1$ and $\mathcal{C}_2$ which are $P_{XV_1V_2}$-covering. Then, for any typical sequence $\mathbf{x}^n$, one can find sequences $\mathbf{c}_i^n\in \mathcal{C}_i, i\in \{1,2\}$ which are typical with $\mathbf{x}^n$ with respect to $P_{XV_1V_2}$. From the Markov Lemma \cite{Markov}, $\mathbf{x}^n$ is typical with $\alpha\mathbf{c}_1^n+\beta\mathbf{c}_2^n$ with respect to $P_{X(\alpha V_1+\beta V_2)}$ since $\alpha V_1+\beta V_2\leftrightarrow V_1,V_2 \leftrightarrow X$. So, by the converse source coding theorem, $\frac{1}{n}\log |\alpha\mathcal{C}_1+\beta\mathcal{C}_2|\geq \log q- H(\alpha V_1+\beta V_2)$ which gives (\ref{NLC4}). The following lemma characterizes the covering bounds for a pair of NQLC's.

\begin{Lemma}
\label{lem:2Qcov}
For any $P_{XV_1V_2}$ on $\mathsf{X}\times\mathbb{F}_q\times \mathbb{F}_q$, parameters $m, n, k_1, k_2, \dotsb, k_m$ and random vectors $(U_{1,i})_{i\in[1,m]}, (U_{2,i})_{i\in [1,m]}$ 
satisfying (\ref{NQLC1})-(\ref{NQLC4}), there exists a sequence of pairs of NQLC's $\mathcal{C}^n_1$ and $\mathcal{C}^n_2$ which are $P_{XV_1V_2}$-covering.
\begin{align}
&\sum_{i\in [1,m]} \frac{k_i}{n}H(U_{1,i}) \geq \log q - H(V_1|X)\label{NQLC1}\\
&\sum_{i\in [1,m]} \frac{k_i}{n}H(U_{2,i}) \geq \log q - H(V_2|X)\label{NQLC2}\\
&\sum_{i\in [1,m]} \frac{k_i}{n}\big(H(U_{1,i})+H(U_{2,i})\big)\geq 2\log q -H(V_1,V_2|X)\label{NQLC3}\\
&\sum_{i\in [1,m]} \frac{k_i}{n}H(\alpha U_{1,i}+\beta U_{2,i}) \geq \log{q}- H(\alpha V_1+ \beta V_2|X),\label{NQLC4}\\
&\qquad\qquad\qquad\qquad\qquad\qquad\qquad\qquad \nonumber \forall {\alpha,\beta\in \mathbb{F}_q\backslash{\{0\}}}.
\end{align}
\end{Lemma}
\begin{IEEEproof}
 Let $X$ be a discrete memoryless source, for typical sequence $x$ with respect to $P_X$, define the following:

\begin{align*}
 \theta (x)&= \sum_{u \in C_1, v\in C_2 } \mathbbm{1}\{(\mathbf{v}_1,\mathbf{v}_2)\in A_\epsilon^n(V_1,V_2|\mathbf{x})\}\\
 & =\sum_{\substack{\mathbf{u}_{1,i},\mathbf{u}_{2,i}\in Z_q^{k_i},\\ i\in[1,m] }}\sum_{(\mathbf{v}_1,\mathbf{v}_2)\in A_\epsilon^n(V_1,V_2|\mathbf{x})} \mathbbm{1}\{\sum_{i\in [1,m]}\mathbf{u}_{1,i}G_i+\mathbf{b}_1=\mathbf{v}_1, \sum_{i\in [1,m]}\mathbf{u}_{2,i}G_i+\mathbf{b}_2=\mathbf{v}_2\}
\end{align*}
Here, $\mathbf{G}_i,\mathbf{b}_1$, and $\mathbf{b}_2$ are chosen randomly and uniformly. For $\mathbf{u}_{1,i}\in \mathbb{Z}_q^{k_i}$, define $g(\mathbf{u}_{1,1}, \mathbf{u}_{1,2},\dotsb, \mathbf{u}_{1,m} )\triangleq \sum_{i\in [1,m]}\mathbf{u}_{1,i}G_i+\mathbf{b}_1$. Similarly define $g(\mathbf{u}_{2,1}, \mathbf{u}_{2,2},\dotsb, \mathbf{u}_{2,m} )\triangleq \sum_{i\in [1,m]}\mathbf{u}_{2,i}G_i+\mathbf{b}_2$ for $\mathbf{u}_{2,i}\in \mathbb{Z}_q^{k_i}$.  

\begin{Lemma}\label{lem: indep} The following hold:
\begin{enumerate}
\item $g(\mathbf{u}_{1,1}, \mathbf{u}_{1,2},\dotsb, \mathbf{u}_{1,m} )$ and $g'(\mathbf{u}_{2,1}, \mathbf{u}_{2,2},\dotsb, \mathbf{u}_{2,m} )$  are uniform over $\mathbb{F}_q^n$.
\item$g(\mathbf{u}_{1,1}, \mathbf{u}_{1,2},\dotsb, \mathbf{u}_{1,m} )$  is independent of $g(\mathbf{\tilde{u}}_{1,1}, \mathbf{\tilde{u}}_{1,2},\dotsb, \mathbf{\tilde{u}}_{1,m} )$  when $(\mathbf{u}_{1,i})_{i\in[1,m]}\neq (\mathbf{\tilde{u}}_{1,i})_{i\in[1,m]}$.
\item $g'(\mathbf{u}_{2,1}, \mathbf{u}_{2,2},\dotsb, \mathbf{u}_{2,m} )$  is independent of $g'(\mathbf{\tilde{u}}_{2,1}, \mathbf{\tilde{u}}_{2,2},\dotsb, \mathbf{\tilde{u}}_{2,m} )$  when $(\mathbf{u}_{2,i})_{i\in[1,m]}\neq (\mathbf{\tilde{u}}_{2,i})_{i\in[1,m]}$.
\item If $\mathbf{b}_1$ and $\mathbf{b}_2$ independent uniform over $\mathbb{F}_q^n$, then $g(\mathbf{u}_{1,1}, \mathbf{u}_{1,2},\dotsb, \mathbf{u}_{1,m} )$  and $g'(\mathbf{u}_{2,1}, \mathbf{u}_{2,2},\dotsb, \mathbf{u}_{2,m} )$ are independent.
\end{enumerate}
\end{Lemma}

\begin{IEEEproof}
Similar to the proof of the covering lemma in \cite{Lattices}.
\end{IEEEproof}

We want to use the Chebyshev's inequality to obtain:

\begin{align*}
P\{ \theta(\mathbf{x})=0 \} &\leq \frac{4var\{ \theta(\mathbf{x})\}}{\mathbb{E}\{\theta(\mathbf{x}) \}^2}\to 0\\
\end{align*}

We calculate the expected value of $\theta(\mathbf{x})$:
  
\begin{align}\nonumber
\mathbb{E}\{ \theta(\mathbf{x})\}&= \sum _{\mathbf{x} \in A_\epsilon^n(X)} \sum_{\substack{\mathbf{\mathbf{u}_{1,i}} \neq \mathbf{u}_{2,i},\\i\in[1,m]}} \sum_{(\mathbf{v}_1,\mathbf{v}_2)\in A_\epsilon^n(V_1,V_2|\mathbf{x})} P(\mathbf{x}) P\{g(\mathbf{u}_{1,1}, \mathbf{u}_{1,2},\dotsb, \mathbf{u}_{1,m} )= \mathbf{v}_1, g'(\mathbf{u}_{2,1}, \mathbf{u}_{2,2},\dotsb, \mathbf{u}_{2,m} )=\mathbf{v}_2\}\\\nonumber
&= \sum _{\mathbf{x} \in A(X)} \sum_{\mathbf{u}_{1,i} \neq \mathbf{u}_{2,i}} | A_\epsilon^n(V_1,V_2|\mathbf{x})| P(\mathbf{x}) \frac{1}{q^{2n}}\\
\label{equ:exp}
&= 2^{n (-\sum_{i\in [1,m]} \frac{k_i}{n}\big(H(U_{1,i})+H(U_{2,i})\big)+H(V_1,V_2|X)+O(\epsilon))}\\\nonumber
\end{align}

%
%
%

The following lemma bounds $\frac{var\{ \theta(\mathbf{x})\}}{\mathbb{E}\{\theta(\mathbf{x}) \}^2}$.
\begin{Lemma}\label{lem: var_exp}
\begin{align*}
&\frac{var\{ \theta(\mathbf{x})\}}{\mathbb{E}\{\theta(\mathbf{x}) \}^2}\\
 &\leq 2^{-n (-\sum_{i\in [1,m]} \frac{k_i}{n}\big(H(U_{1,i})+H(U_{2,i})\big)+H(V_1,V_2|X))} +2^{-n(-\sum_{i\in [1,m]} \frac{k_i}{n}H(U_{1,i})+H(V_1|X))} +2^{-n(-\sum_{i\in [1,m]} \frac{k_i}{n}H(U_{2,i})+H(V_2|X))}\\
&+\sum_{\alpha\in \mathbb{F}_q\backslash\{0\}} 2^{n(-\sum_{i\in [1,m]} \frac{k_i}{n} H(U_{1,i}+\alpha U_{2,i})+H(V_1,V_2|X)-{H(V_1,V_2|X, V_1+\alpha V_2)})}\\
\end{align*}
\end{Lemma}

The proof of the lemma follows from arguments similar to the ones used in \cite{Lattices}. Setting the above to go to 0, we get the covering bounds mentioned in Lemma \ref{lem:2Qcov}

\end{IEEEproof}

\begin{Remark}
 Inequalities (\ref{NQLC1})-(\ref{NQLC3}) are exactly the same bounds on the codebook rates as in (\ref{NLC1})-(\ref{NLC3})   (note that $\sum_{i\in [1,m]} \frac{k_i}{n}H(U_{j,i})$ is the rate of $\mathcal{C}_j, j\in\{1,2\}$.). (\ref{NQLC4}) can also be written as $\frac{1}{n}\log |\alpha\mathcal{C}_1+\beta\mathcal{C}_2|\geq \log q- H(\alpha V_1+\beta V_2)$. By the same argument as in the previous lemma, the bounds can not be tightened by a finer error analysis. The main difference between inequality $(\ref{NQLC4})$ and (\ref{NLC4}) is that in the new bound, the LHS changes as a function of $\alpha$ and $\beta$. This provides new degrees of freedom which in turn result in improvements in the MD problem as shown in the next section. 
 \end{Remark}
\section{Gains in the MD Problem}
\label{sec:MD}
\begin{figure}[!h]
\centering
\includegraphics[height=1.8in]{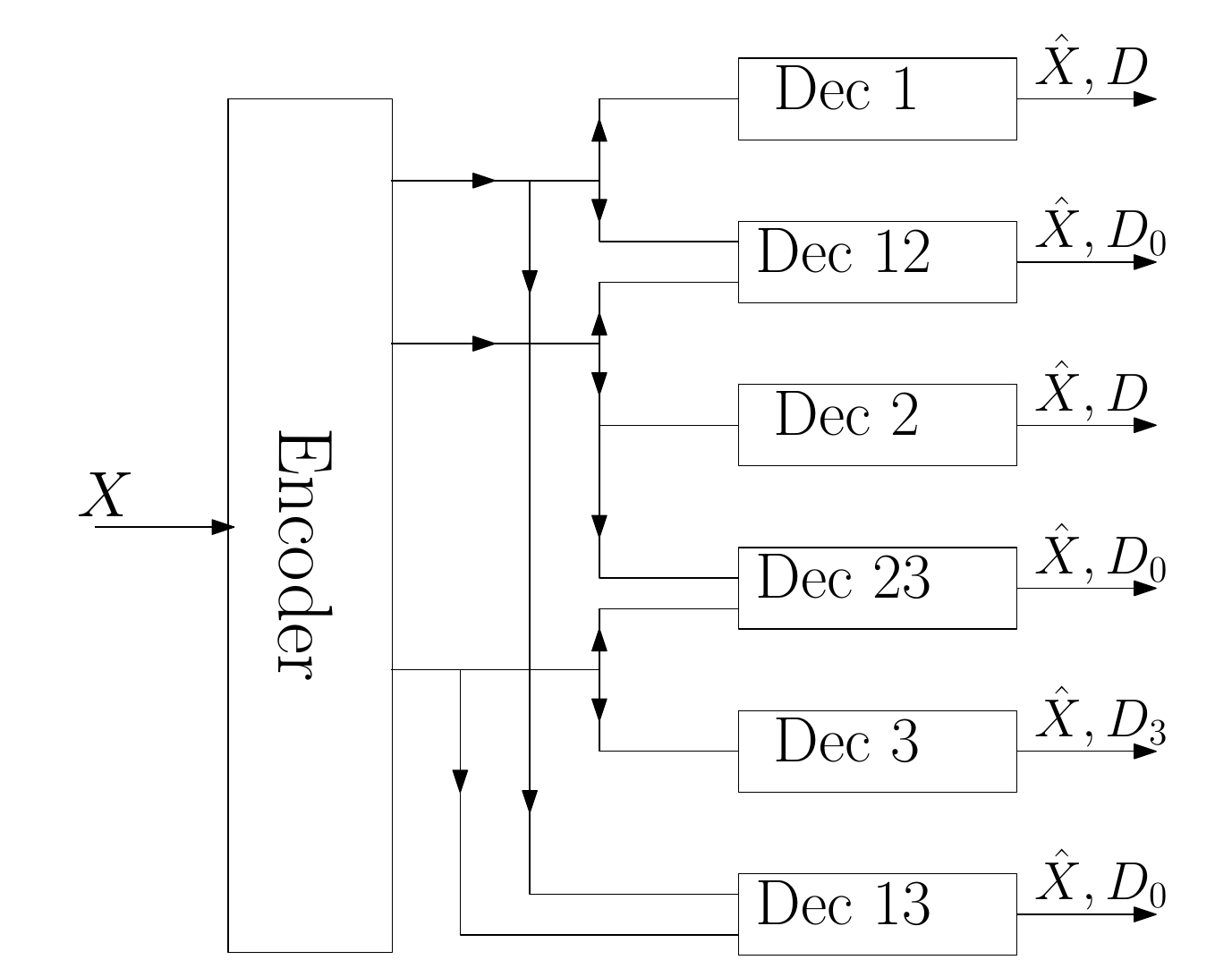}
\caption{A three-descriptions example where NQLC's give gains}
\label{fig_sim}
\end{figure}

In this section, we first present an example in which a scheme based on NQLC's gives improvements in terms of achievable RD's compared to the SSC scheme. The example is constructed by slightly altering example 6 in \cite{MD}. The setup is depicted in Figure \ref{fig_sim}. Here, $X$ is a binary symmetric source. The distortion constraints at all decoders are binary Hamming distortions except for decoder $\{3\}$. Assume that the distortion constraint at decoder $\{3\}$ is such that it needs to reconstruct the ternary addition $\hat{X}_1\oplus_3 2\hat{X}_2$, where $\hat{X}_i, i\in \{1,2\}$ are the reconstructions at decoders $\{1\}$ and $\{2\}$.
We are interested in achieving the RD vectors with the following projections:
%
\vspace{-.1in}
\begin{align}
&R_1=R_2=\frac{1-h_b(D_0)}{2}, D_{\{1\}}=D_{\{2\}}= \frac{1}{2}(1-(1-2D_0)(2-\sqrt{2})),\\
& D_{\{1,2\}}=D_{\{1,3\}}=D_{\{2,3\}}=D_0,\label{RDscalar}
\end{align}
Our objective is to minimize $R_3$ subject to these constraints. The following lemma gives the RD vector achievable using NQLC's which is not present in the RD region in \cite{MD}.

\begin{Lemma}\label{lem:example}
  There exists $\epsilon>0$, such that the RD vector in (\ref{RDscalar}) is achievable using NQLC's, if the following hold:
 \begin{align}
 &R_3\geq H(V_1\oplus_3 2V_2)-H(V_1\oplus_3 V_2|X)-\epsilon\\
 & h_b\left(D_0\right)+2h_b\left(\frac{\sqrt{2}}{2}\right)+h_b\left(2\left(\sqrt{2}-1\right)D_0\right)\nonumber\\&\qquad\qquad\qquad +h_b\left(2\left(\sqrt{2}-1\right)\left(1-D_0\right)\right)=1,
 \label{eq:nonredconstraint}
\end{align}
\label{scalartheorem}
where the joint distribution between $X,V_1$ and $V_2$ is given in table I.
\setlength{\extrarowheight}{.3cm}
\begin{table}[h!]
\centering
\begin{tabular}{x{.9cm}|x{1.2cm}|x{1.54cm}|x{1.5cm}|x{1.7cm}|}
\diag{.1em}{.9cm}{$\qquad X$}{$V_{1},V_{2}$}& \multicolumn{1}{c}{00}&\multicolumn{1}{c}{01}&\multicolumn{1}{c}{10}&\multicolumn{1}{c}{11}\\\hline
0&$\frac{1}{2}(1-D_0)$&$\frac{\sqrt{2}-1}{2}D_0$&$\frac{\sqrt{2}-1}{2}D_0$&$\frac{3-2\sqrt{2}}{2}D_0$\\  \cline{2-5}
1&$\frac{1}{2}D_0$&$\frac{\sqrt{2}-1}{2}(1-D_0)$&$\frac{\sqrt{2}-1}{2}(1-D_0)$&$\frac{3-2\sqrt{2}}{2}(1-D_0)$\\  \cline{2-5}
\end{tabular}
\label{tb:jointdist}
\vspace{0.1in}
\caption{ }
\end{table}
\vspace{-0.1in}

Furthermore, the RD vector is not achievable using the linear coding scheme stated in \cite{MD}.
\end{Lemma}
\begin{IEEEproof}
We provide a scheme which achieves the RD vector for $\epsilon=10^{-4}$ using NQLC's. Let $n$ be large and $\lambda$ a small positive number. construct a pair of $P_{XV_1V_2}$-covering NQLC's with parameters $m=2$, $\frac{k_1}{n}=0.8$, and $\frac{k_2}{n}=0. 2665$ where $U_1$ and $U'_1$ are ternary and uniform. $U_2$ and $U'_2$ have the following distributions,

\begin{table}[h!]
\centering
\begin{tabular}{x{1cm}|x{1.5cm}|x{1.5cm}|x{1.5cm}|}
& \multicolumn{1}{c}{0}&\multicolumn{1}{c}{1}&\multicolumn{1}{c}{2}\\\hline
$U_1$&$0.33$&$0.48$&$0.19$\\  \cline{2-4}
\end{tabular}
\vspace{0.1in}
\end{table}
\vspace{-0.3in}

\begin{table}[h!]
\label{table:dist}
\centering
\begin{tabular}{x{1cm}|x{1.5cm}|x{1.5cm}|x{1.5cm}|}
& \multicolumn{1}{c}{0}&\multicolumn{1}{c}{1}&\multicolumn{1}{c}{2}\\\hline
$U_2$&$0.33$&$0.19$&$0.48$\\  \cline{2-4}
\end{tabular}
\end{table}

Given the above parameters, it is straightforward to check that the constraints in Lemma \ref{lem:2Qcov} are satisfied. Description 1 carries the bin index of $\mathcal{C}$ with bin size $\log{3}-\frac{H(V_1,V_2)}{2}-\lambda$, also, description 2 carries the index for $\mathcal{C}'$ with the same bin size. Description 3 carries the index for $\mathcal{C}\oplus_3 2\mathcal{C}'$ with bin size $\log{3}-{H(V_1\oplus 2V_2)}-\lambda$. Then, 
\begin{align*}
& R_1=R_2=\frac{k_1}{n}+\frac{k_2}{n}H(U_1)-(\log{3}-\frac{H(V_1,V_2)}{2}-\lambda)\\
& R_3=\frac{k_1}{n}+\frac{k_2}{n}H(U_1\oplus_3 2U_2)-(\log{3}-{H(V_1\oplus 2V_2)}-\lambda).
\end{align*}
Direct calculation shows that the above rates are equal to the ones stated in the lemma. We provide an outline of the proof that the scheme in \cite{MD} can not achieve these rates. By the same arguments as in the proof of Example 6 in \cite{MD}, it can be shown that the only non-redundant codebooks in the scheme are $\mathcal{C}_{\{1\}}$, $\mathcal{C}_{\{2\}}$, and $\mathcal{C}_{o,\{3\}}$ (this follows from optimality at decoders $\{1,2\}$ and $\{3\}$, and the uniqueness of the optimizing distribution at decoder $\{1,2\}$ shown in \cite{MD}). Then, in order to satisfy the constraints at decoder  $\{3\}$, we need to set $V_{o,\{3\}}=V_{1}+2V_{2}$. Checking the bounds in \cite{MD} it can be seen that the above rates are not achievable.
.
\end{IEEEproof}

Next, we provide a new achievable RD region for the general MD problem using NQLC's. For brevity, we have only considered the case where a summation of two codebooks decoded at decoders $\{1\}$ and $\{2\}$ is to be decoded at decoder $\{3\}$. So, to achieve the RD region we use all of the codebooks present in the SSC scheme with the addition of a pair of NQLC's. One of the NQLC's is decoded at $\{1\}$, the other at $\{2\}$, and a linear combination of the two is decoded at $\{3\}$ as was the case in the previous example.
 This RD region could be improved upon by considering the reconstruction of an arbitrary number of summations of arbitrary lengths at the decoders as done for the NLC's in \cite{MD}. The notation used in the next definition is the same as in \cite{MD}.

\begin{Definition}
 Fix the prime number $q$. For a joint distribution $P$ on random variables  $U_{\mathcal{M}},\mathcal{M}\in \mathbf{S}_\mathsf{L}$, $V_{\{j\}}, j\in\{1,2\}$, and $X$, where the underlying alphabet for all auxiliary random variables is the field $\mathbb{F}_q$, and a set of reconstruction functions $g_{\mathcal{L}}=\{g_\mathsf{N}:\mathsf{U}_{\mathsf{N}}\to\mathsf{X}, \mathsf{N}\in\mathcal{L}\}$, the set $\mathcal{RD}(P, g_\mathcal{L}) $ is defined as the set of RD vectors satisfying the following bounds for some non-negative real numbers  $(\rho_{\mathcal{M},i},r_{o,\mathcal{M}})_{i\in \widetilde{\mathcal{M}},\mathcal{M}\in {\mathbf{S}_\mathsf{L}} }$, $\rho_{ o,\{j\},i},  i\in \{1,2,3\}$, and parameters $(m,n,k_1,k_2,\dotsb,k_m)$ and vectors of random variables $(A_{i,j})_{j\in[1,m]},i\in\{1,2\}$: 
\begin{align}
& H(U_{\mathbf{M}}V_{\mathbf{E}}|X)\geq   \sum_{\mathcal{M} \in \mathbf{M}}{(\log q  \!  -  r_{o,\mathcal{M}})}+\sum_{\mathcal{E} \in \mathbf{E}}{(\log q  \!  -  r_{o,\mathcal{E}})}, \label{sec1cov1}\\
& H(U_\mathbf{M} , W_{3, \alpha, \beta}|X)\geq   \sum_{\mathcal{M} \in \mathbf{M}}{(\log q  \!  -  r_{o,\mathcal{M}})}+\log{q}-r_{o,3,\alpha,\beta}\label{sec1cov2}\\
& H([U,V,W]_{\overline{{\mathbf{M}}}_\mathsf{N}}|[U,V,W]_{\overline{\mathbf{M}}_\mathsf{N}\cup \overline{\mathbf{L}}})\leq \nonumber\\
&\qquad\!\!\!\!\!\sum_{\mathcal{M}\in {\overline{\mathbf{M}}}_\mathsf{N}\backslash\overline{\mathbf{L}}}\!\!\!\! (\log q +\!\! \sum_{j\in \widetilde{\mathcal{M}}} \rho_{\mathcal{M},j}-r_{o,\mathcal{M}})+\!\!\!\!\!\!\!\!\!\!\!\!\sum_{\substack{\mathcal{M}\in\{\mathcal{A}_1,\mathcal{A}_2,\mathcal{A}_3\}\bigcap \overline{\mathbf{M}}_\mathsf{N},\\j\in \widetilde{\mathcal{M}}}}\!\!\!\!\!\!\!\! \rho_{o,\mathcal{M},j}, \label{sec1pack1}
\\&R_i=\sum_{\mathcal{M}} \rho_{\mathcal{M},i}, \qquad D_{\mathsf{N}}=E\big\{d_{\mathsf{N}}(h_{\mathsf{N}}(U_{\mathsf{N}},X))\big\}.\label{sec1RD}
\end{align}
Where  (a) $\overline{{\mathbf{M}}}_\mathsf{N}\triangleq({\mathbf{M}}_\mathsf{N}, \{ \{j\}|\{j\}\in {\mathbf{M}}_\mathsf{N} \},\big\{\{\{3\}, 1,1\}| \{3\}\in {\mathbf{M}}_\mathsf{N}\big\})$, (b) $\widetilde{\mathbf{M}}_\mathsf{N}\triangleq \bigcup_{\mathsf{N}'\subsetneq \mathsf{N}}{\overline{\mathbf{M}}}_{\mathsf{N}'}$, (c) $r_{o,\{3\},\alpha,\beta}\triangleq \sum_{j\in [1,m]}\frac{k_i}{n}H(\alpha V_{1,j}+\beta V_{2,j})
$, (d) $r_{o,\mathcal{M}}\leq \log{q}$, and (e) $W_{\{3\}, \alpha, \beta}\triangleq\alpha V_{ \{1\}}+\beta V_{\{2\}}$, and the bounds should hold for all $\mathbf{M}\subset \mathbf{S}_\mathsf{L}, \mathbf{E}\subset \{\{1\},\{2\}\}$ and $ \overline{\mathbf{L}}\subset {\overline{\mathbf{M}}}_\mathsf{N}$.
\end{Definition}
 The main difference between this scheme and the one in \cite{MD} is that the rate $r_{o,\{3\},\alpha,\beta}$ is now defined according to the size of the linear combination of NQLC's rather than NLC's.
 
\begin{thm}
  RD vectors in $cl\left(\mathcal{RD}(P, g_\mathcal{L})\right) $  are achievable. Where $cl(\mathsf{A})$ is the closure of set $\mathsf{A}$.
\end{thm}
\begin{IEEEproof}

Given a joint distribution $P_{\mathbf{U},\mathbf{V},X}$, and codebook and binning rates satisfying the bounds in the theorem we prove achievability of the RD vector in (\ref{sec1RD}).

\noindent\textbf{Codebook Generation:} Fix blocklength $n$. For every $\mathcal{M}\in \mathbf{S}_\mathsf{L}$, independently generate a linear code $C_{\mathcal{M}}$ with size $2^{nr_{o,\mathcal{M}}}$. Also generate a pair of NQLC's $C_{ \{j\}}, j\in \{1,2\}$ with parameters as in Definition \ref{def:NQLC} and random variables $(V_{1,j})®$ and $(V_{2,j}), j\in[1,m]$, respectively . Define the set of codewords $C_{o, \{3\},\alpha,\beta}\triangleq   \alpha C_{o,\{1\}}+\beta C_{o,\{2\}}$. The size of $C_{o, \{3\},\alpha,\beta}$ is $2^{nr_{o, \{3\},\alpha,\beta}}$ where $r_{o,\{3\},\alpha,
\beta}= \sum_{j\in [1,m]}\frac{k_i}{n}H(\alpha V_{1,j}+\beta V_{2,j})
$. For the $i$th description bin the codebook $C_{\mathcal{M}}$ randomly and uniformly with rate $2^{n\rho_{\mathcal{M},i}}$ . 

\noindent\textbf{Encoding:} Upon receiving the source vector $X^n$, the encoder finds a jointly-typical set of codewords $c_{\mathcal{M}}$. Each description carries the \textit{bin-indices} of all of the corresponding codewords. The encoder declares an error if there is no jointly typical set of codewords available. 

\noindent\textbf{Decoding:} Having received the bin-indices from descriptions $i\in \mathsf{N}$, decoder $\mathsf{N}$ tries to find a set of jointly typical codewords $c_\mathcal{M}, \mathcal{M}\in \overline{\mathbf{M}}_\mathsf{N}$.  If the set of codewords is not unique, the decoder declares error.

In order for the encoder to find a set of jointly typical codewords, the mutual covering bounds (\ref{sec1cov1}) and (\ref{sec1cov2}) should hold. This is a generalization of the result in lemma \ref{lem:2Qcov} and we omit the proof for brevity. The bounds in (\ref{sec1pack1}) are the mutual packing bounds at each decoder.

\end{IEEEproof}

\section{Conclusion}
A new category of structured codes called QLC's was introduced. The QLC's are constructed by taking specific subsets of linear codes. The tradeoff between application of structured codes and unstructured codes was investigated. We showed that by loosening the linear closure property in linear codes, the mutual covering bounds improve. This improvement led to an MD example where we extracted gains over the previous known coding schemes for the problem. A new achievable RD region for the general MD problem was presented. 
\label{sec:conc}

\end{document}